\documentstyle{mn}
\ifnfsstwo

\fi
 
\ifnfssone
  \newmathalphabet{\mathit}
    \addtoversion{normal}{\mathit}{cmr}{m}{it}
    \addtoversion{bold}{\mathit}{cmr}{bx}{it}

\fi
 
\ifoldfss

\fi
 
\loadboldmathitalic
\loadboldgreek
 
\def\etal{{\it et~al.}~}			% et al. in italics
\def\cf{{{\it cf.},\/} }			% cf. in italics
\title{Topology of COBE Microwave Background Fluctuations}

\author[Colley, Gott \& Park]
	{ Wesley N. Colley $^{1,3}$,
	J. Richard Gott, III, $^1$  and
	Changbom Park $^2$ \\
	$^1$Department of Astrophysical Sciences, Princeton University,
Princeton, NJ 08544-1001 {\tt email: wes, jrg@astro.princeton.edu}\\
	$^2$Department of Astronomy, Seoul National University, Seoul, 151
Korea {\tt email: cbp@astrogate.snu.ac.kr}\\
	$^3$Supported by the Fannie and John Hertz Foundation, Livermore, CA 94551-5032}

\begin{document}
\maketitle

\begin{abstract}
We have measured the topology (genus) of the fluctuations in the cosmic
microwave background seen in the recently completed (four-year) data set
produced by the COBE satellite.  We find that the genus is
consistent with that expected from a random-phase Gaussian distribution, as
might be produced naturally in inflationary models.
\end{abstract}

\begin{keywords}
cosmic microwave background -- cosmology: obseravtions -- large-scale structure
of the universe -- methods: statistical
\end{keywords}

\section{Introduction}
The microwave background radiation (hereafter MBR) discovered by Penzias and
Wilson (1965), has a spectrum close to that of a blackbody with mean
temperature $T = 2.726\pm 0.01K (2\sigma)$ (Mather \etal 1993), with small
scale fluctuations of order $\Delta T/T \sim 10^{-5}$ at $\geq 10 ^\circ$ as
observed by COBE (Smoot \etal 1992), in agreement with inflationary models with
various values of $\Omega$ (\cf Gott 1982, 1986; Bardeen, Steinhardt, and
Turner 1983; Bond 1988; Vittorio and Juszkiewicz 1987).  Recently,
the final (four-year) data have been reported by COBE on the World Wide Web
(COBE 1996).

The geometric nature of the MBR anisotropy may be assessed by mapping
isotemperature contours.  Local geometrical information about structures in the
map is contained in three quantities: total area above (or below) a threshold,
total contour length and curvature of the contour (\cf Willmore 1982).  These
are local and invariant quantities of the contours in the sense that they can
be calculated even from a temperature map with incomplete and patchy coverage
and do not change under translation and rotation of the coordinate frame.  At
high threshold levels the excursion regions of a random temperature field will
appear as isolated hot spots surrounded by a cold background and the total
curvature will be positive.  At a low-temperature threshold the contours
surround cold spots and the total curvature is negative.  Near the mean
temperature the hot and the cold regions are nearly symmetric, and the total
curvature is close to zero.

We define the genus $G$ of the excursion set for a random temperature field on
a plane as

\[
\begin{array}{ll}
G = &\mbox{(number of isolated high-temperature regions)}- \\
~&   \mbox{(number of isolated low-temperature regions)}\\
\end{array}
\]
Equivalently, the genus can be defined as the total curvature of the contours.
Assuming a contour defines a differentiable curve $C$ on the map, its total
curvature is given by the integral

\[
K = \int_C \kappa ds \equiv 2\pi G
\]
where $\kappa$ is the local curvature, $s$ parameterizes the curve, and $G$ is
the genus of a single contour.

A hot spot will contribute $+1$ to the total map genus and a cold spot
(``hole'') in it will decrease the genus by 1.  Therefore the genus can be
considered as the total number of connected hot regions minus the total number
of holes in them.  In practice, contours may cross the edge of the survey
region, in which case the partial curves contribute non-integer rotation
indices to the genus.

According to (Gott \etal 1990), a two-dimensional random-phase Gaussian
temperature field will generate a genus per unit area

\[
g \propto \nu e^{-\nu^2/2}
\]
where $\nu$ is the threshold value, above which a fraction, $f$ of the area
has a higher temperature

\[f = (2\pi)^{-1/2}
\int_\nu^\infty \exp(-x^2/2)dx\]
(\cf also Adler 1981; Melott \etal 1989; Coles 1988; Gott \etal 1992; Park
\etal 1992).

\section{COBE Observations}
For our analysis we have used the linear combination of the 6 COBE
(four-year) DMR maps provided by the COBE team, which properly subtracts the
Galaxy while minimizing the noise (COBE Archive 1996), (for method, see Bennett
\etal 1992).  This map is then smoothed with a Gaussian window function with a
FWHM of $7^\circ$.  Since the COBE antennas have a FWHM beam width of $7^\circ$
this amounts to an effective Gaussian smoothing window function with FWHM of
$10^\circ$ for any real signal.  To eliminate any residual galactic
contamination at low galactic latitude we will only consider the North and
South galactic caps with $|b| > 30^\circ$.

\begin{figure}
\vspace*{224pt}
\caption{Total genus (North$-$South) is plotted as a function of $\nu$.  The
size of the error-bars has been estimated from the observed {\em rms} of all
(North$-$South) genus differences at fixed $\nu$.  The best-fit theoretical
curve for a Gaussian random-phase distribution ($g \propto \exp(\nu^2/2)$) is
shown as a solid line.  The temperature-scale is given at the top.}
\end{figure}

The genus of the temperature anisotropies is plotted as a function of $\nu$ in
figure 1.  Across the top of the figure the temperature scale is shown---that
it is nearly linear shows that the temperature histogram in the smoothed map is
approximately Gaussian.  The best-fit theoretical genus curve expected for a
random-phase Gaussian distribution: $g(\nu)\propto\nu\exp(-\nu^2/2)$ (eq. 3) is
shown as a solid line.  The total genus (North$+$South) is plotted, and a
single estimate of the size of the error bars (in the total genus) is made from
the {\em rms} of all the observed (North$-$South) genus differences at fixed
$\nu$.  As figure 1 shows, the observed genus curve is consistent with the
random-phase theoretical curve within the errors (thus confirming similar
findings from the COBE one-year data [Torres 1993; Smoot
\etal 1994; Park and Gott 1993; Kogut 1993]).

\section{Discussion}
For comparison, we have constructed and analyzed simulated pure noise maps
using COBE error estimates at each position on the sky, which gives $G(\nu=1) =
48 \pm 4$.  Our observed best-fit value for $G(\nu=1)$ = 31 (figure 1) is over
four standard deviations below that expected for a pure noise map.  This
independently confirms the Smoot \etal 1992 conclusion that we are observing
real signal here and not just receiver noise.

If the real signal is Gaussian random-phase, then addition of instrumental
noise (which is itself presumably locally Gaussian random-phase) should leave
us with a genus curve which approximately obeys the Gaussian random-phase law,
$g \propto \exp(\nu^2/2)$ as seen in figure 1.  Since the signal-to-noise ratio
here is approximately 0.4, only gross deviations from Gaussian random-phase in
the real signal could possibly reveal themselves in the figure.  Nonetheless,
the agreement between the data and the random-phase theory shown in figure 1
demonstrates conclusively that the topology of the MBR from the COBE results
does not falsify the Gaussian random-phase hypothesis.

We have presented here the genus measurement of temperature anisotropies in the
microwave background as measured by the COBE satellite and found it to be
consistent with the inflationary prediction of Gaussian random-phase
anisotropies.

\section{Acknowledgements}
The COBE datasets were developed by the NASA
Goddard Space Flight Center under the guidance of the COBE Science Working
Group and were provided by the NSSDC.  WNC thanks the Fannie and John Hertz
Foundation for its continued and gracious support.  This research has been
supported by NASA Grant NAG5-2759.

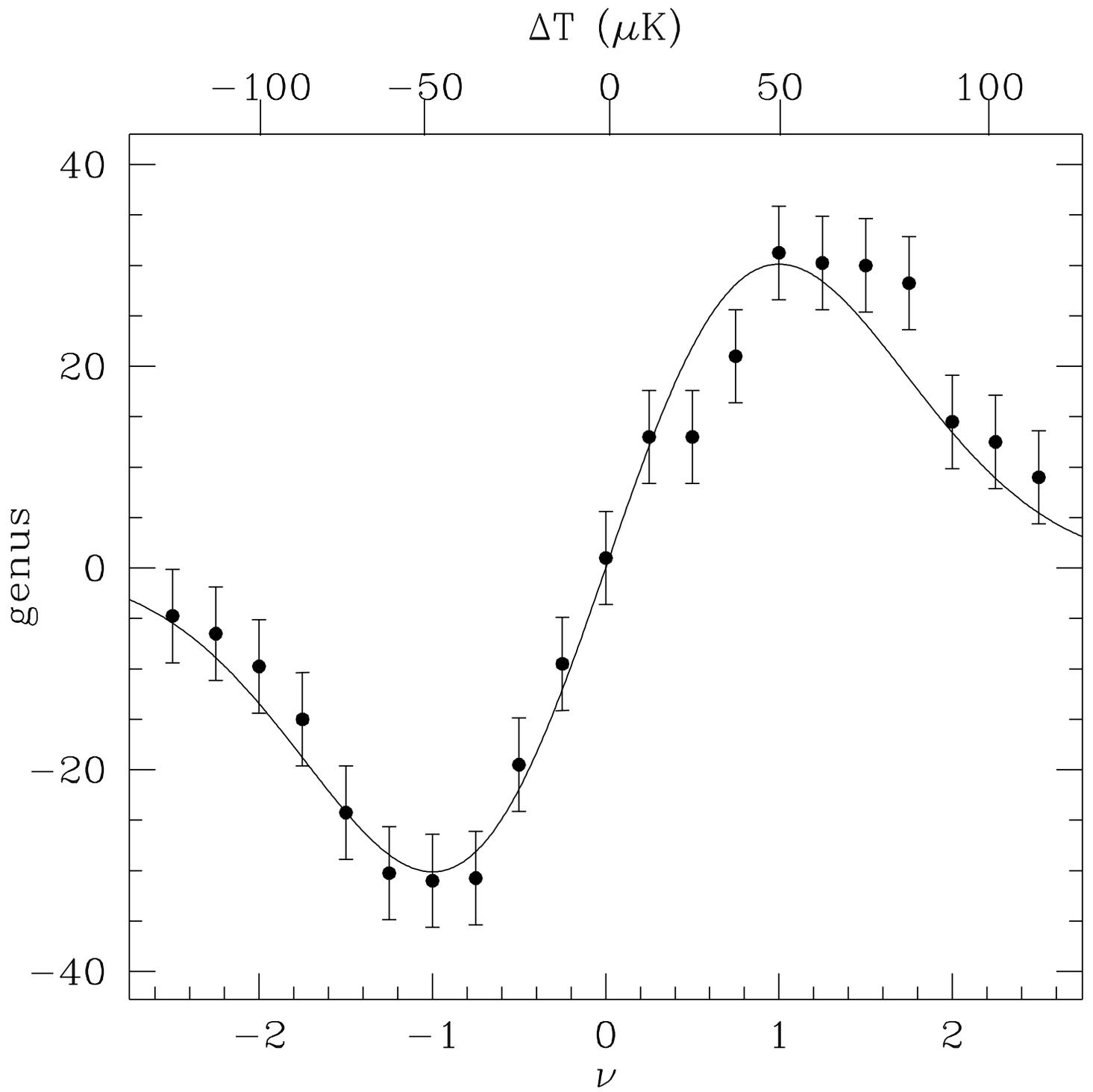
\end{document}